\documentclass[aps,twocolumn,prd,preprintnumbers, 10pt]{revtex4-1}

\usepackage{amsmath,amsfonts,amssymb}
\usepackage{bm}
\usepackage[utf8]{inputenc}
\usepackage{lmodern}
\usepackage{mathrsfs}
\usepackage{mathtools}
\usepackage{tensor}
\usepackage{amsthm}
\usepackage{tikz-cd}

\theoremstyle{definition}

\usepackage{tikz}
\usetikzlibrary{arrows,decorations.markings}
\tikzset{
  big blue arrow/.style={
    decoration={markings,mark=at position 1 with {\arrow[scale=2,#1]{>}}},
    postaction={decorate},
    shorten >=0.4pt},
  big blue arrow/.default=blue}
  \tikzset{
  big red arrow/.style={
    decoration={markings,mark=at position 1 with {\arrow[scale=2,#1]{>}}},
    postaction={decorate},
    shorten >=0.4pt},
  big red arrow/.default=red}
    \tikzset{
  big arrow/.style={
    decoration={markings,mark=at position 1 with {\arrow[scale=2,#1]{>}}},
    postaction={decorate},
    shorten >=0.4pt},
  big arrow/.default=black}

\newcommand\bR{\mathbb{R}}
\newcommand\bZ{\mathbb{Z}}

\newcommand{\nn}{\nonumber}

\newcommand{\nc}{\newcommand}
\nc{\rnc}{\renewcommand} 

\rnc{\a}{\alpha}
\rnc{\b}{\beta}
\nc{\g}{\gamma}
\rnc{\l}{\lambda}
\rnc{\d}{\delta}
\nc{\e}{\epsilon}
\nc{\ee}{\varepsilon}
\nc{\z}{\zeta}
\nc{\f}{\phi}
\nc{\m}{\mu}
\nc{\n}{\nu}
\rnc{\r}{\rho}
\rnc{\k}{\kappa}
\rnc{\l}{\lambda}
\nc{\s}{\sigma}
\rnc{\t}{\tau}
\nc{\w}{\omega}
\nc{\x}{\chi}
\nc{\F}{\Phi}
\rnc{\L}{\Lambda}
\usepackage{pgfplots}
\pgfplotsset{compat=newest}
\pgfplotsset{colormap={violet}{rgb255=(25,25,122) rgb255=(238,140,238) color=(white)}}

\usepackage[hidelinks]{hyperref}

\usepackage{comment}
\usepackage[draft,inline,nomargin,index]{fixme}
\FXRegisterAuthor{kr}{akr}{\color{orange}KR}

\begin{document}

\title{Symmetry Operators and Gravity}
\author{Ibrahima Bah}
\author{Patrick Jefferson}
\author{Konstantinos Roumpedakis}
\author{Thomas Waddleton}

\affiliation{William H. Miller III Department of Physics and Astronomy, Johns Hopkins University,\\ 3400 North
Charles Street, Baltimore, MD 21218, U.S.A.
}

\begin{abstract}
    We argue that topological operators for continuous symmetries written in terms of currents need regularization, which effectively gives them a small but finite width. The regulated operator is a finite tension object which fluctuates. In the zero-width limit these fluctuations freeze, recovering the properties of a topological operator. When gravity is turned on, the zero-width limit becomes ill-defined, thereby prohibiting the existence of topological operators. 
\end{abstract}

\maketitle
\tableofcontents
\section{Introduction}
In recent studies, symmetry has taken center stage in many different and seemingly unrelated arenas of physics. This stems from the novel observation that symmetry structure is encoded in the properties of topological operators, i.e., operators which commute with the energy-momentum tensor, and hence for which correlation functions are insensitive to small deformations of their position \cite{Gaiotto:2014kfa}.
Excitingly, this generalization of symmetry brings to bear a rich mathematical structure, dubbed categorical symmetries, in the study of quantum systems.  The implications of these developments have injected new life in the study of fundamental physics, with important consequences yet to be uncovered.

On the other hand, it is widely expected that UV complete theories of gravity cannot admit any conserved charges \cite{Hawking:1976ra,Banks:2010zn,Coleman:1988cy,Giddings:1988cx,Abbott:1989jw,Coleman:1989zu,Kallosh:1995hi,Arkani-Hamed:2006emk,Harlow:2018jwu,Harlow:2018tng}.  This leads to a curious paradox involving quantum field theories -- which can emerge as low-energy limits of gravitational theories -- and quantum gravity.  This paradox can be further articulated with the following question: what is the fate of conservation laws in field theories when embedded in gravity from a bottom-up perspective, and how do they emerge from the top-down perspective in decoupling limits of gravity?  In the context of generalized symmetry, this question can be formulated as: what is the fate of topological operators in the presence of gravity?  Note that here, we wish to study the fate of operators measuring charge instead of charge conservation.

In this paper, we provide a concrete perspective on how to address this question and set the stage for analyzing its broader implications in gravity.  

Quantum field theories often admit a wide class of extended objects, such as solitons, domain walls, defects, and topological operators to name a few.  We can study their role in QFT by their explicit insertion in the path integral.  Often, this is done at the level of the fields by considering special boundary conditions, or interfaces where the fields undergo non-trivial transformations.  This perspective is usually taken in the study of solitonic objects, which are finite energy configurations.  From the point of view of fields, one can study the role of the collective coordinates in the world-volume dynamics of such solitonic objects \cite{Tong:2005un}.  A particular class of collective coordinates that is always present corresponds to the Goldstone modes associated to the spontaneous breaking of spacetime symmetries due to the insertion. These are most easily understood in terms of the explicit choices of boundary conditions imposed on the fields.  These modes constitute a universal sector of the world-volume dynamics of solitons, and therefore are integral to any effective field theory description of these objects.  

\begin{figure}[t]
    \centering
    \begin{tikzpicture}
        \draw (-3,1.4) -- (-3,-1.8);
        \draw (-3,1.4) -- (-1.2,1.8);
        \draw (-3,-1.8) -- (-1.2,-1.4);
        \draw  (-1.2,1.8) -- (-1.2,-1.4);
        \node at (-0.6,0) {$\rightarrow$};
        \node at (1,0) {\includegraphics[scale = 0.18]{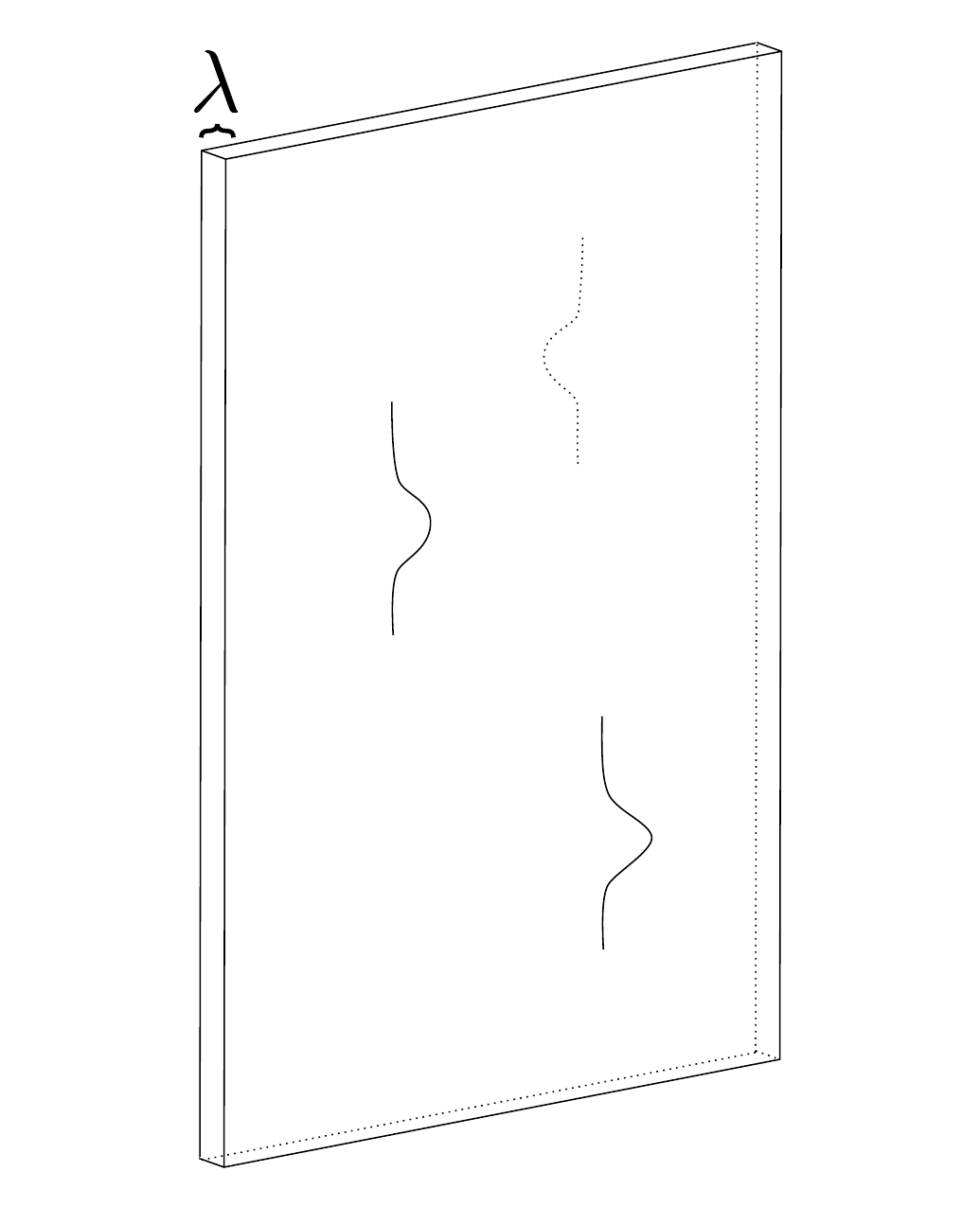}};
    \end{tikzpicture}
    \caption{A regulated topological defect acquires a non-zero width $\lambda$ and fluctuates.}
    \label{fig:Defect}
\end{figure}
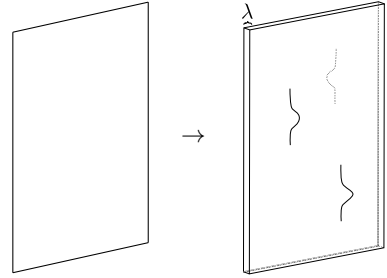

In string theory, a quantum theory of gravity, there are studies of symmetries that have shed light on how topological operators can emerge in the decoupling limit of gravity. For example, in \cite{GarciaEtxebarria:2022vzq,Apruzzi:2022rei} it was demonstrated that the topological sector of branes in string theory (i.e., the Wess-Zumino terms in the low energy effective worldvolume action) yield topological operators in field theories constructed both using geometric engineering and AdS/CFT duality.  This approach to studying branes as topological operators was further explored in \cite{Bah:2023ymy,Apruzzi:2023uma,Heckman:2022peq,Heckman:2022xgu,Cvetic:2023pgm,Heckman:2024oot,Waddleton:2024iiv}.  The main lesson from these works is that extended non-topological defects in the string theory can become topological operators in the decoupling limit of gravity. In particular, the kinetic terms of the collective coordinates of the branes freeze when $G_{\rm N}$ is taken to zero.  

This top down perspective inspires our main approach in this paper, as it suggests that we can analyze topological operators when coupling field theories to gravity by treating them as solitons.  While it would be interesting to provide a purely field-theoretic justification for this assumption, we settle with this motivation from string theory.  With this assumption, our aim is to determine whether or not we can see the collective coordinates come alive in a universal fashion when gravity is turned on.  Said differently, our goal is to demonstrate that there is a universal sector, associated to Goldstone modes of spontaneously broken spacetime symmetries, that is necessarily dynamical in presence of gravity.  

It is also worth pointing out that in a theory of gravity, the geometry (and presumably also the topology) of the spacetime background fluctuates.  We therefore expect the dynamics of the collective coordinates to be induced by these fluctuations, and furthermore that the dynamics of the collective coordinates freeze or decouple in the field theory limit. 

In this paper, we wish to study the insertion of topological operators similarly as solitons \footnote{Throughout this paper we use the terms ``operator'' and ``defect'' interchangeably, since we are working with QFTs that are defined in Euclidean space.}.  This is achieved by considering field configurations whose asymptotic values are related by a symmetry transformation, as depicted in Figure \ref{fig:Defect}.  The ``soliton" interpolates between the two field configurations, has a finite width and supports an effective field theory for the collective coordinates.  We will characterize the limit wherein the ``soliton" recovers a topological operator. We expect decoupling or freezing of the collective modes in such a limit.  An important difference between solitons and topological operators is that the former are finite energy configurations, while the latter have vanishing stress tensor.

The above solitonic picture regulates the insertion of the operator. We first discuss why a choice of regulator is needed in the case of continuous symmetries with Noether currents (see \cite{Cherman:2021nox} for a related discussion).  Once this point is established, we assume a definition of topological operators in field theory as solitons in general, including those that generate finite symmetries.

This approach to defining topological operators will set the stage for how we study their fate in gravity.  In particular, our objective will be to demonstrate that the collective coordinates cannot freeze or decouple when $G_{\rm N}$ is small but non-vanishing.



In Section \ref{sec:Operator_Insertion}, we explore our prescription for defining topological operators in some examples and understand how the collective coordinates freeze. We consider simple scalar field theories with 0-form symmetries, and the electric 1-form symmetry in Maxwell theory. We then couple these theories to gravity in Section \ref{sec:gravity} and demonstrate how the topological limit is ill-defined.
\section{Inserting a Symmetry Defect}\label{sec:Operator_Insertion}

In this section, we argue that a topological defect $U(\Sigma)$ implementing a continuous symmetry transformation needs to be regulated and, in doing so, acquires a small but non-zero width $\lambda$. This is equivalent to a large but finite tension $T \sim \frac{1}{\lambda}$. The topological defect is then recovered in the infinite tension limit $T \to \infty$ (equivalently, $\lambda \to 0$). This sets the stage for exploring the fate of such operators when $G_N$ is non-vanishing.  

To illustrate the argument, we begin with a co-dimension one defect $U(\Sigma)$, generating a continuous 0-form symmetry. To keep the notation simple, consider a $U(1)$ symmetry associated to a classically conserved current $d*j=0$. The symmetry action is generated by the topological operator

\begin{equation}
\label{UsualOp}
    U_\alpha(\Sigma) = e^{i\a \oint_\Sigma *j }~.
\end{equation}
This is a composite operator and needs regularization for reasons we explain below \footnote{For local operators the analogous procedure is point-splitting.}.

Let us insert this defect inside a correlation function. For concreteness,
consider a complex scalar field $\phi$ with action given by
\begin{equation}
    S[\phi] = \int d^dx \;\left(\partial^\m \phi^\dagger \partial_\m \phi - V(|\phi|^2)\right)~.
\end{equation}
This model has a $U(1)$ symmetry $\phi \rightarrow e^{i\theta }\phi$ associated to the current

\begin{equation}
\label{U(1)-current}
    j_\m = i(\phi^\dagger \partial_\m \phi - \phi \partial_\m \phi^\dagger)~.
\end{equation}
We take $\Sigma$ to be a flat surface along the constant time slice $t=0$. We would like to evaluate a correlation function of the form

\begin{align}
\label{correlator}
    &\left< \phi(t_1, \Vec{x}_1)\cdots e^{i\a \int_{t=0} d^{d-1} x \; j_0 } \phi(t_i, \Vec{x}_i)\cdots\right>~,
\end{align}
where $t_1, \dots, t_{i-1} < 0 < t_i,\dots$.

To evaluate the correlation function \eqref{correlator} by means of the path integral, we can use a field redefinition that absorbs the exponential in \eqref{correlator} and multiplies the fields on one side of the defect by phases. 
A naive field redefinition to consider is 
\begin{equation}
    \phi \rightarrow  e^{i\a \; \Theta(t)} \phi~, \label{FRedeF}
\end{equation}
where $\Theta(t)$ is the Heaviside step function. The existence of this field redefinition implies that $U_\alpha(\Sigma)$ is topological. 

In the path integral that computes the above correlation function, namely
\begin{equation}
    \int D\phi \; e^{-S[\phi]} \phi(t_1, \Vec{x}_1)\cdots e^{i\a \int_{t=0} d^{d-1} x \; j_0 } \phi(t_i, \Vec{x}_i)\cdots,
\end{equation}
the transformation of the kinetic term produces a term that cancels with the defect. The transformation also correctly modifies
only the fields inserted at times $t>0$.
However, this redefinition also produces an ill-defined singular term  proportional to $\delta(t)^2$ where $\delta(t)$ is a delta function. Thus, to proceed, we need regularize the field redefinition \eqref{FRedeF} so that ill-defined $\delta(t)^2$ does not appear; we accomplish this by replacing the step function with a smooth regulator function \footnote{The class of regulator functions we consider are called sigmoid functions.}.

Although the result does not depend on the details of the regulator function, for concreteness let us choose the function
\begin{equation}
    f_{\lambda}( t) = \frac{1}{1+e^{- t/\lambda}}~.
\end{equation}
This is a one-parameter family of functions parameterized by $\lambda$ such that
\begin{equation}
    \lim_{\lambda\to 0}f_{\lambda}( t) = \Theta(t)~.
\end{equation}
The limit $\lambda\rightarrow 0$ is the zero-width limit of the defect.

We use the function $f_\lambda(t)$ to regulate the defect by making the replacement
\begin{equation}\label{RegDef}
    e^{i\a \int_{t=0} d^{d-1} x  \; j_0 } \rightarrow e^{ \int d^{d}x \left( i\a f'_{\lambda} (t) \; j_{0} - \a^2 f'_{\lambda}(t)^2 |\phi|^2\right)} ~,
\end{equation}
where we also added a counterterm which blows up in the $\lambda\rightarrow 0$ limit. This counterterm is needed so that the field redefinition 
\begin{equation}
\label{fieldredef}
    \phi \rightarrow  e^{i\a \, f_{\lambda}(t)} \phi~,
\end{equation}
absorbs the defect and transforms the fields with the above phases \footnote{Note that the Jacobian determinant associated to the spacetime-dependent field redefinition of the complex scalar in (\ref{fieldredef}) is trivial.}. After the field redefinition, we can take the limit $\lambda\rightarrow 0$ and arrive at the expected result, namely the fact that $U_\alpha(\Sigma)$ is topological and can be freely moved around in correlation functions.

One might worry the addition of the counterterm might destroy the topological nature of $U_\alpha(\Sigma)$. However, the insertion of the defect \eqref{RegDef} corresponds to turning on the flat background gauge field
\begin{equation}
    A = f_{\lambda}'(t) dt~,
\end{equation}
for the $U(1)$ global symmetry. It is straightforward to calculate the divergence of the energy-momentum tensor in this background, for which we find
\begin{equation}
    \partial^\m T_{\m\n} = F_{\n \m}\left(i(D^\m\phi)^\dagger\phi - i\phi^\dagger D^\m\phi\right) ~,
\end{equation}
where $F=dA$ is the field strength and $D= d + iA$ the covariant derivative. Hence, we see that as long as the background gauge field is flat, i.e., $F=0$, the energy-momentum tensor is conserved, which implies that \eqref{RegDef} is topological. This also illustrates the necessity for adding the counterterm in \eqref{RegDef}, coming from the $A^2 |\phi|^2$ term in the kinetic term.

Another way to derive the counterterms in \eqref{RegDef} is through the Ward-Takahashi identities, which are the quantum version of the classical conservation equations $\partial_\m j^\m=0$ and $\partial^\m T_{\m \n}=0$.
In the case of the complex scalar, the Ward-Takahashi identity reads
\begin{align}
\label{WTid}
    \left< \partial_\m j^\m(x) \Psi(x_1) \cdots\right>
    = \sum_i q_i \delta(x-x_i)\left< \Psi(x_1)  \cdots\right> ~,
\end{align}
where $\Psi$ is either $\phi$ or $\phi^\dagger$ and $q_i = \pm$ is the charge of $\Psi(x_i)$. If we integrate by parts and rewrite the operator in \eqref{RegDef} as

\begin{equation}
\label{RegDef2}
    U_\alpha = e^{ \int d^{d}x \left( - i\a f_{\lambda}(t) \; \partial_\m j^\m - \a^2 f'_{\lambda}(t)^2 |\phi|^2\right)}~,
\end{equation}
then after expanding the exponential, the term linear in $\alpha$ can be evaluated using the Ward-Takahashi identity above. To evaluate the quadratic term we need the Ward-Takahashi identity with two current insertions. Starting with \eqref{WTid} and using the expression for the current in \eqref{U(1)-current}, one can derive the identity
\begin{widetext}
\begin{align}
    \left< \partial_\m j^\m(x) \partial'_\n j^\n(x') \Psi(x_1) \cdots\right>-2
    \partial'_\m\left(\partial^\m\delta(x-x')\left< |\phi(x')|^2 \Psi(x_1) \cdots\right>\right)=
     \sum_{ij} q_iq_j \delta(x-x_i)\delta(x'-x_j)\left< \Psi(x_1)  \cdots\right> \nn~.
\end{align}
\end{widetext}
In order to get the expected result, namely the charge of field-insertions after integrating the divergences of the currents as in \eqref{RegDef2}, we need to subtract the second term on the left-hand side above, and this is ensured by the presence of the term $|\phi|^2$  in \eqref{RegDef2}.

To summarize, to define the defect $U_\alpha(\Sigma)$ we gave it a small but non-zero width. This causes the field to transform smoothly across the defect. Motivated by this picture as well as the top down perspective in holography, we are going to view a symmetry defect as a solitonic object which can fluctuate around its classical position. 
The insertion of this solitonic object spontaneously breaks spacetime symmetry,  and the fluctuations correspond to the dynamics of the resulting Goldstone excitations living on the soliton---these will be the modes of interest when studying its dynamics \cite{Low:2001bw}.

As we have stressed above, in order for a topological operator to be well-defined in a given quantum field theory, it must be possible to take the zero-width limit $\lambda \to 0$, and in doing so, any Goldstone modes living on the defect worldvolume must necessarily either freeze or decouple. In what follows of this section, we explore our prescription for defining topological operators in some simple examples, and for each of these examples, we argue that the Goldstone modes are frozen in the zero-width limit.
\subsection*{Shift symmetry}

Let's consider a free scalar field $\phi(x)$ in $d$-dimensional flat spacetime with action 
\begin{equation}\label{eq:shift_action}
    S = \frac{1}{2} \int{ d}^d x\; \partial_\mu \phi ~\partial^\mu \phi~. 
\end{equation}

The theory is invariant under the shift transformation $\phi \to \phi + \alpha$ and the associated Noether current is 
\begin{equation}\label{eq:shift_current}
    \qquad j_\mu = \alpha~ \partial_\mu \phi~. 
\end{equation}

We can construct a solitonic model of this transformation as follows. To start, let $\Sigma$ be a codimension-1 interface at $t=t_0$. We consider a field profile that smoothly interpolates between the original field configuration as $t\to -\infty$ and the transformed configuration as $t\to +\infty$. That is, we consider the profile
\begin{equation}\label{eq:shift_field_redef}
    \phi(\sigma,t) \rightarrow \phi(\sigma,t) + \alpha\,f_{\lambda}(t-t_0)~,
\end{equation}
where $\{\sigma^a\}$ denote the coordinates of $\Sigma$. With the field profile above we see that in the limit $\lambda\to 0$ the value of $\phi$ transforms discontinuously across the interface located at $t = t_0$, thereby giving the desired behavior of a symmetry operator. 

So far, our regulated topological shift symmetry defect behaves as a finite width object. In the bulk quantum theory, such an object fluctuates around its classical position at $t=t_0$. We can model these fluctuations by promoting $t_0 \to t_0(\sigma^a)$ to a collective coordinate that depends on the worldvolume coordinates of the defect; this mode can be interpreted as the Goldstone boson for the broken translation symmetry along the direction normal to the defect. In the rest of this section we analyze the effective theory of the Goldstone modes on the defect.

The effective action should comprise of both universal and non-universal terms, where the latter depend on the specifics of the defect at hand. The diffeomorphism-invariant universal term that contains the most relevant kinetic terms is the Nambu-Goto action 

\begin{equation}
    T \int_\Sigma d^{d-1} \sigma \sqrt{ \det \frac{\partial x^\m}{\partial \s^a}  \frac{\partial x_\m}{\partial \s^b}}~,
\end{equation}
where $T$ is the bare tension. In our setup, the defect's classical position is the flat plane located at $t=0$. Expanding about small fluctuations, the defect effective action becomes 
\begin{equation}
    \label{NG}
     \frac{T}{2}  \int_\Sigma d^{d-1} \sigma (1 + (\partial_a t_0)^2 + \cdots)~,
\end{equation}
where the dots include terms with higher powers of $t_0$. 
Substituting \eqref{eq:shift_field_redef} into \eqref{eq:shift_action}, 
we find that the effective action takes the form
\begin{align}
    \begin{split}\label{eq:shift_new_lag}
        S = &\int{ d}^d x \;  \Big[\frac{1}{2}(\partial_\mu \phi)(\partial^\mu\phi) + \alpha\,f_{\lambda}'\left(\partial_t\phi - \partial^a\phi\partial_at_0\right) \\ 
        &+\frac{\alpha^2}{2 }f_\lambda'^2\left( 1 + (\partial_a t_0)^2 \right)  {\Big ]}~ .
    \end{split}
\end{align}

In the zero-width limit $\lambda \rightarrow 0$, the coefficient of the second term approaches a delta function $\delta(t)$; this implies that when integrating over the $t$ direction, we can restrict the integrands to their values at $t = 0$, which localizes the resulting action to the worldvolume of the interface. The third term, which on the defect yields a term proportional to $\frac{1}{\lambda}$, is a correction to the bare tension in \eqref{NG} \footnote{The regulator we are using has the property $f_\lambda'^{2} = \frac{1}{6 \lambda} f'_\lambda - \frac{\lambda}{6}f'''_\lambda$}.

The above considerations imply that when $\lambda$ is taken to be sufficiently small, the action can be approximated as
	\begin{align}
	\label{eq:approx_shift_action}
		S &= S_{\text{bulk}} + S_{\rm d} + S_{\rm G}+ \mathcal \cdots\nn ~,\\
		S_{\rm G} &= \int d^{d-1}\sigma\left[\frac{T_{\rm eff}}{2}(\partial_a t_0)^2  -\alpha\partial^a{\phi}\partial_at_0 \right] ~,
  \end{align}
where $S_\text{bulk}$ is the bulk action in \eqref{eq:shift_action}, $S_{\rm d}$ is the regulated defect as in \eqref{RegDef} and $S_{\rm G}$ is the effective action for the Goldstone mode. We have also introduced the effective tension

\begin{equation}
    T_{\rm eff} = T +\frac{\a^2}{6\lambda}~.
\end{equation}

In the $\lambda\to 0$ limit, the path integral over $t_0$ can be treated with a saddle point or semiclassical approximation, which freezes the collective coordinate $t_0$ to be a constant. The remaining integral over the zero mode leads to an infinite constant which we can absorb in the normalization factor of the defect. Hence, in the zero-width limit we land on the regulated version \eqref{RegDef2} of the topological operator.

\subsection*{Phase shift of complex scalar}
Let us now go back to the complex scalar field $\phi$ with a $U(1)$ symmetry $\phi \rightarrow e^{i\theta }\phi$ and take $\Sigma$ to be a flat surface along a constant time slice $t=0$. Following a similar analysis as in the previous section we arrive at
\begin{align}
     S_{\rm G} =\int d^{d-1} \sigma \left[ \frac{T}{2}(\partial ^at_0)^2 +i\a \partial ^at_0 \; j_a -\frac{\alpha^2}{12 \lambda} |\phi|^2 (\partial ^at_0)^2\right] \nn~.
\end{align}
The first term is the kinetic term for the Goldstone as well as interaction terms with the bulk.  As in the case of the shift symmetry, in the limit $\lambda\to 0$  the Goldstone must freeze for arbitrary field profile of $\phi$ in order to recover the topological operator.   
\subsection*{Higher-form symmetry}
The previous construction can be generalized to incorporate more general $p$-form symmetries. Recall that for such a symmetry, its generators are described by operators $U_\alpha(\Sigma)$, defined on codimension-$(p+1)$ submanifolds of spacetime and labeled by an $\alpha$ parameterizing the symmetry group. Let us consider the operators charged under this symmetry to be represented by holonomies of $p$-form gauge fields $C_p$. If the charged objects are denoted $W_q(M^p)$ where $q$ indicates the charge, then we can compute the action of our symmetry in correlation functions similar to \eqref{correlator}.
\begin{equation}
    \left\langle U_\alpha(\Sigma) W_q(M)\cdots \right\rangle = e^{2\pi i q\alpha\,\ell(\Sigma,M)}\left\langle W_q(M)\cdots\right\rangle,
\end{equation}
where $\ell(\Sigma,M)$ denotes the linking number of $\Sigma$ and $M^p$ in the ambient spacetime \cite{Bott1982}. We can view this action equivalently as the action of a shift of $C_p$ by a flat $p$-form $\Lambda_p$ with periods $\int\Lambda_p = \alpha$ valued in the symmetry group \footnote{For $p$-form symmetries with $p>0$, the symmetry group is necessarily Abelian, so we can consider the symmetry group to be $U(1)$, a finite subgroup, or products of these.}. In this description, by inserting an operator $U_\alpha(\Sigma)$ that links with $W_q(M^p)$, we are adding a source to the field strength as
\begin{equation}\label{eq:gen_higher_sym_field_configuration}
    G_{p+1} = \hat{G}_{p+1} + \alpha\, \delta(\Sigma)~,
\end{equation}
where $G_{p+1}$ is the field strength of $C_p$, $\hat{G}_{p+1}$ is the field strength away from the symmetry defect, $\alpha$ is a constant valued in the symmetry group, and $\delta(\Sigma)$ is the Poincar\'e dual of $\Sigma$ in the ambient spacetime. Once we regulate this field profile and introduce the Goldstone modes, wherein we will find one mode for each direction transverse to $U_\alpha(\Sigma)$, the exact expression for $\delta(\Sigma)$ will be dependent on the regulating functions $f_\lambda(t^i-t^i_0)$ and hence the Goldstone modes $t^i_0$. Note that the example of the shift symmetry earlier gives a realization of this transformation for $p=0$.

As a more concrete example, let us consider free Maxwell theory in flat 4d spacetime described by the following action,
\begin{equation}\label{eq:Maxwell_action}
    S = \int d^4 x\left(\frac{1}{4e^2}F^{\mu\nu}F_{\mu\nu}\right) ~,
\end{equation}
where $e$ is the usual electromagnetic coupling. This theory has a $U(1)_e$ 1-form symmetry acting on the Wilson lines $W_q(M^1)\to e^{i\alpha a}W_q(M^1)$, which as discussed above can be equivalently described by a shift of the gauge field $A\to A+\Lambda_1$ where $\Lambda_1$ is a flat connection whose periods are valued in $\bR/\bZ$.

We now wish to model this transformation using a soliton constructed as discussed above. We will start with a codimension-2 submanifold $\Sigma^2$ located at the locus $t^1 = t^2 = 0$, where we are using $t^i$ to denote the different directions transverse to $\Sigma$. Our proposed field profile is the application of the previous paragraphs where $p=1$. That is, we take
\begin{equation}
    F \to F+\alpha\,\delta(\Sigma) = \hat{F} + \alpha\,\delta(t^1)\delta(t^2)dt^1dt^2\,.
\end{equation}
We can regulate this profile as before, and find a representative field profile for the gauge field $A$ as follows
\begin{equation}\label{eq:Maxwell_field_redefinition}
    A\to A + \frac{\alpha}{2}\left(f_{\lambda,1}{f'_{\lambda,2}}d(\Delta t^2) - {f'_{\lambda,1}}f_{\lambda,2}d(\Delta t^1)\right)~.
\end{equation}
Here we use $f_{\lambda,i}$ to denote the regulated step function depending on $\Delta t^i \equiv t^i - t_0^i$. By performing a similar analysis as the previous examples, taking care to account for the two transverse directions, we find that the action \eqref{eq:Maxwell_action} with the profile \eqref{eq:Maxwell_field_redefinition} may be approximated in the following way,
\begin{align}\label{eq:approxMaxwellAction}
    \begin{split}
	   S &= S_{\text{bulk}} + S_{\rm d} + S_{\rm G} + \cdots\\
	   S_{\rm G} &= \int d^{2}\sigma\, \Big[\frac{T}{2}(\partial^a t_0^i)^2  + \alpha \partial_a t_{0,i}\, j^{ai}\\
        &\qquad\qquad + \frac{\alpha^2}{72e^2\lambda^2}(\partial_a t_0^i)^2+ \dots\Big],
    \end{split}
\end{align}
where the ellipses in the first line above denote terms higher order in $\lambda$. Above we have only included terms that are quadratic order or below in the Goldstone modes. The terms represented by dots in $S_{\rm G}$ include terms of higher order in the Goldstone modes that reproduce the Nambu-Goto action as well as higher order couplings to the bulk field strength; the latter are independent of $\lambda$. In the zero-width limit we see that each of the individual Goldstone modes freeze out, and the resulting defect is topological as expected.
\section{Gravity}\label{sec:gravity}
In the previous section, we modeled a symmetry defect as a solitonic object with width $\lambda$ and studied the induced EFT for the Goldstone modes living on the defect worldvolume. We then argued that in the limit $\lambda\to 0$ these Goldstone modes freeze and the soliton describes a topological symmetry operator. In this section, we demonstrate that in the presence of gravity, there is an obstruction to taking the zero-width limit and thus having a topological operator. Hence, we arrive at the conclusion that in a gravitational theory, there is no operator measuring a conserved charge.

In general, the effective theory on a codimension-$(p+1)$ defect will be of the form
\begin{equation}\label{eq:universal}
     \int { d}^{d-p-1} \sigma \; \frac{T}{2}  \sqrt{ \det\left(g_{\mu\nu} \frac{\partial x^\m}{\partial \s^a}  \frac{\partial x^\n}{\partial \s^b}\right)}+\cdots ~,
\end{equation}
where we are using $x^\m$ to collectively denote $x^a = \sigma^a$ and $x^i = t^i - t_0^i$ \footnote{This is equivalent to choosing static gauge had we started with the embedding coordinates as our dynamical degrees of freedom.}. The ellipses in the above equation denote higher order terms as well as non-universal interactions between the ``Goldstones'' $t_0^i$ and the bulk fields \footnote{Strictly speaking, in the presence of gravitational interactions (with spacetime symmetry being gauged), these fields are not Goldstone bosons, but rather correspond to longitudinal polarizations of the graviton.}. They may act as corrections to the bare tension $T$ of the defect.

We wish to consider the bulk theory coupled to gravity and analyze the contribution to the effective tension. To this end, we define the graviton field $h_{\m\n}$ by expanding around a fixed background, which we take to be flat for simplicity. The metric takes the form

\begin{equation}\label{eq:graviton}
    g_{\m\n} = \delta_{\m\n} + \kappa h_{\m\n}~,
\end{equation}
where $\kappa^2 = 32\pi^2 G_{\rm N}$ defines a perturbative parameter with $G_{\rm N}$ being Newton's constant. From \eqref{eq:universal}, the effective action contains terms of the following form:
\begin{align}\label{eq:graviton_expansion}
\begin{split}
S &= \int d^{d-p-1}\sigma \;\frac{\kappa\,T}{2}\Big(h_{ij}\partial^at_0^i\partial_at_0^j \\
&\quad + h_{ab}\partial^a t_0^i\partial^b t_{0,i} - 2h_{ai}\partial^at_0^i + h_{a}\!^a \Big) + \cdots ~.
\end{split}
\end{align}
Note in particular the graviton tadpole term--- it dictates the leading coupling of the defect with gravity, and gives the interpretation of the coupling $T$ being a ``tension'' of the defect \cite{Polchinski:1996na}. 
In addition to those in \eqref{eq:graviton_expansion}, there will be non-universal terms from the bulk action giving interactions between the Goldstone modes and the graviton. These additional terms act as various correction terms similar to the non-universal terms in \eqref{eq:universal}.

After performing the linearization \eqref{eq:graviton} and choosing a field profile enacting a transformation, the total action of the theory can be schematically written as
\begin{equation}
    S = S_{\rm bulk}(\kappa) + S_{\rm d}(\lambda) + S_{\rm G}(\lambda) + S_{\rm g}(\lambda,\kappa) ~,
\end{equation}
for $\lambda$ taking a small but non-zero value. The first term, $S_{\rm bulk}(\kappa)$ is the original bulk action, including the standard contributions from graviton interactions in linearized gravity. The second and third factors recover the defect action as derived in the previous sections. The final factor, $S_{\rm g}(\lambda,\kappa)$, consists of all additional terms containing the graviton field that act as corrections to $S_{\rm G}(\lambda)$.

The exact form of $S_{\rm g}(\lambda,\kappa)$ depends on the bulk theory; the relevant terms for our analysis are those correcting the graviton tadpole in \eqref{eq:graviton_expansion} above. The corrections coming from $S_{\rm g}(\lambda,\kappa)$ modify this coupling and allow us to define an effective tension in the worldvolume theory, similar to the previous sections. The effective action of the defect will contain the term
\begin{equation}\label{eq:effective_tadpole}
    \int d^{d-p-1}\sigma\; \frac{\kappa\,T_{\rm eff}(\phi,\lambda)}{2}h_a\!^a~ ,
\end{equation}
where we have made explicit that the effective tension may depend on both the width $\lambda$ as well as the values of bulk fields on the defect worldvolume. As we will see in examples below, the $\lambda$ dependence of the effective tension is such that when the zero-width limit is taken, $T_{\rm eff}$ diverges. Previously we argued that this divergence indicated that the worldvolume modes must freeze in order to have a well-defined defect theory. However, in \eqref{eq:effective_tadpole} such terms are absent, and so we cannot sensibly consider the limit $\lambda\to 0$. This suggests that the presence of the graviton, or equivalently a non-zero value of $\kappa$, acts as an obstruction to taking the zero-width limit. Thus, for $\kappa\neq 0$ we cannot define a topological operator, and we cannot meaningfully measure a conserved charge.

The divergence discussed in the previous paragraph is signaling that the defect is strongly-coupled in the gravitational theory. Physically, this is the statement that the defect is in a regime where it is best described by a black brane. As the width $\lambda$ decreases, the energy density increases and perturbation theory breaks down as the defect experiences stronger gravitational interactions. If we wish to have a well-defined EFT on the defect worldvolume, we naturally find a hierarchy to the parameters in the theory---if we wish to take the zero-width limit, then we must first take the limit $\kappa \to 0$. However, there cannot exist any effective description of the worldvolume dynamics on the defect in the regime
\begin{equation}
    \frac{\kappa}{\lambda} \geq M_{\rm p}^{-\frac{d-4}{2}} ,
\end{equation}
where $M_{\rm p}$ is the Plank mass. This implies that in theories with gravitational interactions, we are stuck with the dynamics of the collective coordinates on the defect. In the remainder of this section we explore the consequences of this observation in the context of the examples we discussed previously.
\subsection*{Examples}
\paragraph{Shift symmetry} Let us first return to the example of a free scalar field in $d$ dimensions with shift symmetry. For convenience, we repeat the action,
\begin{equation}
    S = \frac{1}{2}\int d^d x\,\partial_\mu\phi\partial^\mu\phi\,,
\end{equation}
as well as the field profile that models the shift symmetry of this theory,
\begin{equation}
    \phi(\sigma,t) \rightarrow \phi(\sigma,t) + \alpha\,f_\lambda(t-t_0)~.
\end{equation}
By combining the above with the metric perturbations \eqref{eq:graviton}, we find the correction
\begin{align}
\begin{split}
    \int{ d}^{d-1}\sigma\,\frac{\alpha^2\kappa}{12\lambda}h_a\!^a~,
\end{split}
\end{align}
which then defines the effective tension by
\begin{equation}
    T_{\rm eff}(\phi,\lambda) = T + \frac{\alpha^2}{6\lambda}~.
\end{equation}
From here we see that the effective tension diverges as we take the limit $\lambda\to 0$, causing the defect to become gravitationally strongly-coupled. Therefore, we find an ill-defined worldvolume theory as long as $\kappa$ is non-zero.

\paragraph{Complex scalar} Our next example is that of the complex scalar theory, with action given by
\begin{equation}
    S = \int d^dx\;\left(\partial^\mu\phi^\dagger\partial_\mu\phi - V(|\phi|^2)\right)~.
\end{equation}
The field profile modeling the $U(1)$ symmetry of this theory, as discussed above, is given by
\begin{equation}
    \phi(\sigma,t) \to e^{i\alpha f_\lambda(t-t_0)}\phi(\sigma,t)~.
\end{equation}
Upon expanding the metric to account for small perturbations, there is a contribution to the graviton tadpole term on the defect given by
\begin{equation}
    \int d^{d-1}\sigma\,\frac{\alpha^2\kappa}{6\lambda}|\phi|^2 h_a\!^a~,
\end{equation}
thus allowing us to define an effective tension
\begin{equation}
    T_{\rm eff}(\phi,\lambda) = T + \frac{\alpha^2}{3\lambda}|\phi|^2~.
\end{equation}
We again find that the worldvolume theory fails to be well-defined in the limit $\lambda\to 0$ as long as $\kappa$ is present. Note here that the divergent behavior is sensitive to the bulk field $\phi$, as alluded to in the comment following \eqref{eq:effective_tadpole} above.

\paragraph{Maxwell theory} Our final example consists of 4d Maxwell theory, using the action and field profile described in \eqref{eq:Maxwell_action} and \eqref{eq:Maxwell_field_redefinition}, respectively. Upon considering the effects of metric fluctuations through the field $h_{\mu\nu}$, we find on the defect worldvolume the following graviton tadpole contributions.
\begin{equation}
    \int d^2\sigma\,\frac{\alpha^2\kappa}{72e^2\lambda^2}h_a\!^a~.
\end{equation}
This allows us, as in the previous examples, to define an effective tension,
\begin{equation}
    T_{\rm eff}(\phi,\lambda) = T + \frac{\alpha^2}{36e^2\lambda^2}~.
\end{equation}
The characteristic to note in this case is that the power of $\lambda$ in the denominator has increased due to the higher number of transverse directions to the defect. The tension still diverges as we try to take the zero-width limit, and in fact it diverges faster than the effective tensions in the 0-form symmetry examples.

%
%
%
%
%
%
%
%
%
%
%
%
%
%
%
%
%
%
%
%
%
%
\section{Summary and Outlook}

In this work we have focused on topological operators for continuous symmetries acting on bosonic fields, realizing them as solitonic objects corresponding to smooth profiles for the fundamental fields. A key feature of continuous symmetries is that they admit Noether currents, which provides a mechanism for obtaining the couplings between the bulk theory and the fluctuations of the collective coordinates on the defects. We used the interactions between the collective coordinates and the bulk theory to determine a mechanism by which a defect becomes topological as its width is taken to zero; as we have demonstrated, this automatically imposes the freezing of the associated Goldstone modes. Moreover, we have used these examples to argue that in the presence of gravity, the zero-width limit of such defects is obstructed due to the emergence of strong gravitational effects that indicate the back-reaction of the defect on the bulk theory. This obstruction offers a complementary perspective on the expected absence of symmetries in gravitational theories.  We expect that our general conclusions about the existence of defect dynamics to hold in more generic cases (some of which we discuss below), as well as in cases for which gravitational interactions are present from the beginning of the analysis.  

An intriguing potential implication of our results is that would-be symmetries in QFTs coupled to gravity, without gauging, imply the existence of dynamical defects whose physics is in principle accessible within perturbation theory.  This may lead to new insights into symmetry-violation in gravitational theories. More specifically, the new interactions involving the graviton generate a running of the parameter, $\lambda$. One can explicitly check that $\lambda$ receives no quantum corrections in the field theory examples above for $\kappa=0$. An important task will be to understand this running of $\lambda$ in gravity.  

Our analysis touches on many additional ideas, and there are numerous future directions that we anticipate will bring these ideas further into focus, as well as clarify various subtle issues that have emerged along the way. We comment below on some of these future avenues.

For discrete symmetries no Noether current exists, but there are still Goldstone modes arising from spontaneous breaking of spacetime symmetries. Without the Noether current, there is no obvious coupling between the Goldstones and the bulk theory. It would be interesting to understand how we can see the freezing of the Goldstone modes in the case of discrete symmetries. 

Our methods naively do not yield the corrections to the tension for the Goldstone modes when the symmetries are acting on fermions. This seems to be a technical issue relating to the fact that fermionic theories have one-derivative kinetic terms while bosonic theories have two-derivative kinetic terms. It will be important to analyze topological operators acting on fermionic fields.

Another interesting direction is to study cases of more general categorical symmetries. There are many examples of such symmetries, signified by the presence of a conserved current that is not gauge invariant \cite{Yokokura:2022alv, Choi:2022fgx}. These cases can be incorporated into our analysis, and we plan to explore them in future work. 

Recent work \cite{Apruzzi:2022rei, GarciaEtxebarria:2022vzq, Heckman:2022muc, Heckman:2024oot} has explored the existence of topological operators in holographic theories. These cases have been a strong motivation for analyzing the fate of topological operators in QFTs coupled to gravity in terms of the dynamics of soliton-like objects. In these works, it was demonstrated in the context of AdS/CFT that such objects can become topological on the conformal boundary of AdS, leading to the proposal that branes ``are'' symmetry operators. This is consistent with our discussion since $\kappa$ is renormalized to zero on the boundary. In light of our perturbative field theory analysis, we also expect that there exists a regulated version of an extended dynamical object in AdS that, when pushed to the conformal boundary, becomes a regulated version of a topological operator in the holographic dual field theory. Thus, another natural next step is to carefully work out a suitable version of a regulated worldvolume action for branes in the AdS bulk that is compatible with our field theory analysis when pushed to the conformal boundary. 

Finally, as mentioned above an intriguing potential application of our results is the characterization of symmetry/charge violation. Since we have essentially argued that topological defects, which can be viewed as implementing the measurement of charge, fail to be topological due to the presence of gravitational interactions, we anticipate that the existence of dynamical modes on the defect can be related to lower bounds the uncertainty of charge measurements in the presence of gravity. It would be very interesting and useful to be able to express such a bound in terms of the tension and other parameters that control the  dynamics of the collective coordinates on the defect worldvolume. We defer a thorough investigation of these ideas and applications to future work.

\let\oldaddcontentsline\addcontentsline
\renewcommand{\addcontentsline}[3]{}
\paragraph*{{\bf Acknowledgments}}
\let\addcontentsline\oldaddcontentsline
We are grateful to  Zohar Komargodski for comments on the manuscript. We would like to thank Fabio Apruzzi, Nima Arkani-Hamed, Federico Bonetti, Thomas Dumitrescu, Jonathan Heckman, Simeon Hellerman, Ken Intriligator, David E. Kaplan, Shlomo Razamat, Sakura Schafer-Nameki, Yuji Tachikawa, Cumrun Vafa, and Amos Yarom for illuminating discussions. 
PJ is supported by the Johns Hopkins University Provost's Postdoctoral Fellowship.
IB, KR, PJ, and TW are supported in part by the Simons Collaboration on Global Categorical Symmetries and also by the NSF grant PHY-2412361.

\let\oldaddcontentsline\addcontentsline
\renewcommand{\addcontentsline}[3]{}
\bibliography{references}
\let\addcontentsline\oldaddcontentsline

\end{document}